\documentclass[%
 reprint,
 amsmath,amssymb,
 aps,
]{cup-hpl}
\usepackage{graphicx}
\usepackage{dcolumn}
\usepackage{bm}
\usepackage[separate-uncertainty=true,multi-part-units=single]{siunitx}
\usepackage{amsmath}
\usepackage{booktabs}
\usepackage{leftidx}
\usepackage{lipsum}

\newcommand{\tulio}{\ensuremath{\mathrm{Tm}\mathrm{:Lu}_{2}{\mathrm{O}_3}}}
\newcommand{\ped}[1]{\ensuremath{_\mathrm{#1}}}
\newcommand{\peff}{\ensuremath{P\ped{eff}}}
\newcommand{\pthr}{\ensuremath{P\ped{Thr.}}}
\newcommand{\fref}[1]{Fig.~\ref{#1}}

\begin{document}
\shorttitle{A study of cross-relaxation and temporal dynamics of lasing at 2 microns in Thulium doped ceramic}
\shortauthor{A. Fregosi {\it et al.}}

\title{
A study of cross-relaxation and temporal dynamics of lasing at 2 microns in Thulium doped ceramic
}

\author[1]{Alessandro Fregosi\corresp{Alessandro Fregosi, Fernando Brandi, and Luca U. Labate, Consiglio Nazionale delle Ricerche, Istituto Nazionale di Ottica (CNR-INO), Pisa, Via Moruzzi, 1, Pisa 56124, Italy Emails: alessandro.fregosi@ino.cnr.it (A. Fregosi); fernando.brandi@ino.cnr.it (F. Brandi); luca.labate@ino.cnr.it (L. Labate)}}
\author[1]{Fernando Brandi}
\author[1]{Luca Labate}
\author[1]{Federica Baffigi}
\author[1]{Gianluca Cellamare}
\author[1]{Mohamed Ezzat}
\author[1]{Daniele Palla}
\author[2]{Guido Toci}
\author[1]{Alex Whitehead}
\author[1]{Leonida A. Gizzi}

\address[1]{Intense Laser Irradiation Laboratory (ILIL) CNR-INO, Pisa, Via Moruzzi, 1, Pisa 56124, Italy}
\address[2]{Consiglio Nazionale delle Ricerche, Istituto Nazionale di Ottica (CNR-INO) Via Madonna del Piano 10, Sesto Fiorentino 50019, Italy}

\begin{abstract}
We report the characterization of the pump absorption and emission dynamic properties of a \tulio{} ceramic lasing medium using a three mirrors folded laser cavity.
We measured a slope efficiency of 73\%, which allowed us to retrieve the cross-relaxation coefficient.
The behavior of our system was
modeled via a set of macroscopic rate equations
in both the quasi continuous wave and the pulsed pumping regime.
Numerical solutions were obtained, showing a good agreement with the experimental findings.
The numerical solution also yielded a cross-relaxation coefficient in very good agreement with the measured one, showing that  the cross-relaxation phenomenon approaches the maximum theoretical efficiency.
\end{abstract}

\maketitle

\section{\label{sec:level1}Introduction}
The use of multi-TW ultrashort pulse lasers  has been emerging dramatically in the past decades for fundamental studies and multi disciplinary applications \cite{Danson_2019}. Their effectiveness in exciting and driving plasma waves, for example, makes this class of lasers ideal as drivers of laser-plasma accelerators that are being considered for the next generation of compact light sources and are being investigated for future colliders for high energy particle physics. The laser specifications in terms of repetition rate, and therefore average power, required for these applications are beyond current industrial capabilities, limited to a few tens of Watts, with the most advanced scientific systems now in the 100 W range.  Large, laser-based plasma accelerator infrastructures currently under construction\cite{Assmann2020,10232104,Werle2023} are based on PW-scale peak power lasers, with ultra-short pulse duration, down to 30 fs or less, and an energy per pulse up to 100 J, at a repetition rate for user applications up to 100 Hz and beyond.
These projects rely on laser systems that are mostly based on Ti:Sa, ideally with pump lasers featuring diode pumping. However, the demanding specifications of pump lasers for Ti:Sa, requiring nanosecond pulse duration and relatively short wavelength, limit scalability of this technology.\\
Indeed, the possibility of scaling plasma acceleration further to meet particle physics needs \cite{adli:hal-02050782}, requires much higher efficiency, beyond the capabilities of most established technologies, thus calling for new solutions.
A number of different approaches, based on entirely new concepts, materials and architectures, are being developed to overcome fundamental limitations of present laser systems in terms of wall-plug efficiency, compactness and, ultimately, average power.
Among these novel schemes, those based on Thulium doped materials lasing at  2 micron wavelength have been proposed as a promising ultrashort pulse laser platform with high average power, high repetition rate \cite{10.1117/12.2525380} for their potential high-energy storage capability \cite{Tamer2024_1}, mainly because of the long fluorescence time, of the order of milliseconds, and the convenient pumping wavelength, just below 800 nm. These features enable diode pumping with industrial-grade systems and also operation in the so-called multi-pulse extraction regime \cite{10.1117/12.2520981} at very high repetition rate.
Notably, 2 micron high-power high repetition rate laser systems with nanosecond pulse duration are currently being investigated as promising solid-state sources for improved EUV lithography systems based on laser-driven tin microdroplet plasma emission \cite{Behnke2021,Behnke2023,Mostafa2023}.\\
Recently, short pulse operation of Tm:YLF was also demonstrated \cite{Tamer2024_2} with TW level peak power, confirming the potential of this platform.
Thulium doped polycrystalline ceramic materials are also being considered as gain media due to their high thermal conductivity, scalability, cost-effectiveness and doping flexibility \cite{rastogi_2024}. Among those materials,  ceramic $\mathrm{Tm:Lu}_{2}\mathrm{O}_3$ along with other Thulium doped sesquioxides are being explored for their exceptional thermal conductivity, higher than that of  any other laser material, suitable for relatively thick disk architectures \cite{Palla2022, Cellamare2021}.
In spite of the large quantum defect set by the 2 µm lasing wavelength, Thulium doped materials can exhibit efficient cross-relaxation (CR), a mechanism in which the energy of excitation, initially taken by one ion, is partially transferred to a neighboring ion originally in the electronic ground state, leaving both ions in the upper laser level \cite{1980_Powell}. While cross-relaxation has been observed in Thulium doped materials, the extent to which this mechanism can be exploited remains an open issue, raising the need for a more extensive experimental investigation.

In this paper we investigate the role of cross relaxation in polycrystalline ceramic $\mathrm{Tm:Lu}_{2}\mathrm{O}_3$
with 4~at.\% doping, by considering the detailed steady state dynamics and the accurate modeling of the pump and laser waist in the medium to carefully evaluate the absorbed laser energy. Our experimental results show that in our conditions cross-relaxation is very efficient, with the coefficient approaching 1.9 and leading to a slope efficiency well exceeding 70\%.
\section{\label{sec:model}Theoretical model for the Tm ion emission dynamics}

\begin{figure}[t]
\centering
\includegraphics[width=0.8\columnwidth]{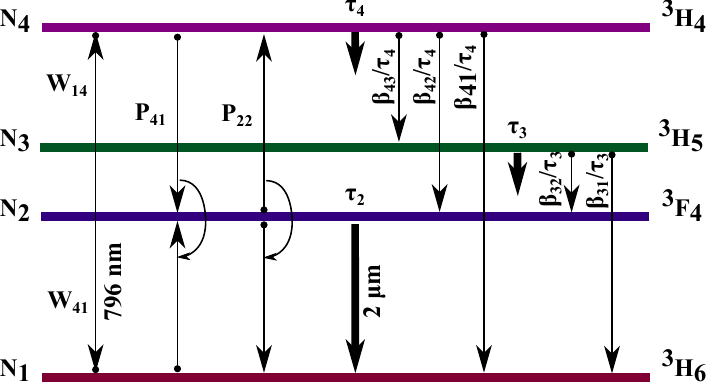}
\caption{The scheme of the energy levels used to model the laser dynamics, from \cite{Albalawi_2017}.}
\label{fig:rateeq}
\end{figure}

\begin{figure}[b]
\centering
\includegraphics[width=0.9\columnwidth]{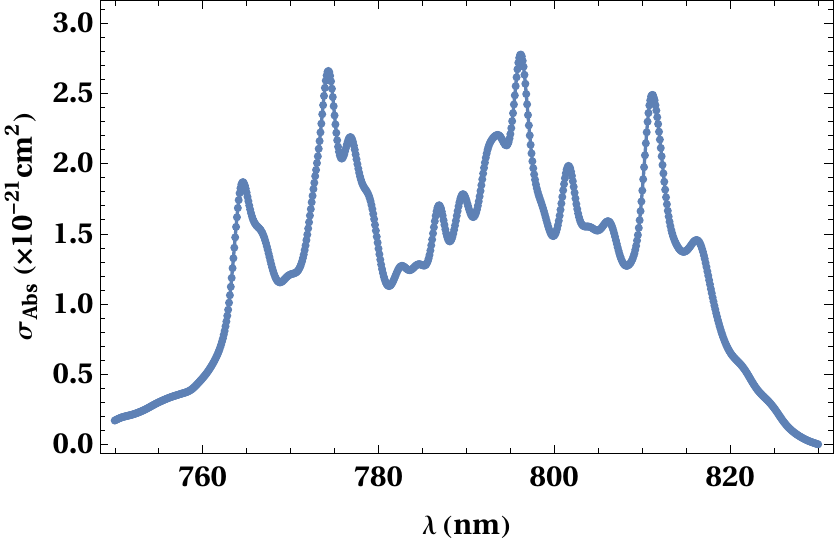}
\caption{Example of the measured absorption cross-section in the center of  the \tulio{} ceramic sample used.}
\label{fig:sigmaabs}
\end{figure}

In order to simulate the steady state dynamics of the \tulio{} ceramic laser, we consider the energy levels and the transitions shown in Fig.(\ref{fig:rateeq}). The rate equations can be obtained from the ones in \cite{Albalawi_2017} as:
\begin{align}
\label{Eq.rate_1}
\frac{dN_4}{dt}&=W_{14}N_1-W_{41}N_4-\frac{N_4}{\tau_4}-P_{41}N_4 N_1+P_{22}N_2^2\\
\label{Eq.rate_2}
   \frac{dN_3}{dt}&=-\frac{N_3}{\tau_3}+\frac{\beta_{43}N_4}{\tau_4}\\
   \nonumber
   \\
\begin{split}
\label{Eq.rate_3}
    \frac{dN_2}{dt}&=2P_{41}N_4 N_1-2P_{22}N_2^2-\frac{N_2}{\tau_2}+ \frac{\beta_{42}N_4}{\tau_4}+\frac{\beta_{32}N_3}{\tau_3}\\&+A_L N_1-E_L N_2
\end{split}\\
\begin{split}
\label{Eq.rate_4}
\frac{dN_1}{dt}&=W_{41}N_4-W_{14}N_1+P_{22 }N_2^2-P_{41} N_4 N_1+\frac{N_2}{\tau_2}\\&+\frac{\beta_{41}N_4}{\tau_4}+\frac{\beta_{31}N_3}{\tau_3}+E_L N_2 -A_L N_1.
\end{split}
\end{align}
where $N_1$, $N_2$, $N_3$ and $N_4$ are the population densities of the levels $^3\mathrm{H_6}$, $^3\mathrm{F_4}$, $^3\mathrm{H_5}$ and $^3\mathrm{H_4}$ respectively.
The sum $N_1+N_2+N_3+N_4=N$ is given by the total ion density, which can be inferred by the doping level. The spontaneous emission lifetime of the \textit{i}-level is given by $\tau_i$, while $\beta_{ij} $ are the $i\rightarrow j$ level branching ratio, with $\sum_j \beta_{ij}=1$. The pump rates are defined by $W_{14}\left (t \right)=\sigma_a I_p \left (t \right )/h \nu_p$ and $W_{41}\left (t \right)=\sigma_e I_p \left (t \right )/h \nu_p$, where $\sigma_a$ and $\sigma_e$ are respectively the absorption and emission pump cross section obtained from the measurements reported in \fref{fig:sigmaabs}, $I_p$ is the pump intensity, \textit{h} is the Planck constant and $\nu_p$ is the frequency of the pump laser. The functions $E_L \left ( t\right )=\sigma_{eL} I_L \left (t \right )/h \nu_L$ and $A_L \left ( t\right )=\sigma_{aL} I_L \left (t \right )/h \nu_L$ specify the lasing rates with $\sigma_{aL}$ and $\sigma_{eL}$ estimated from the data reported in \fref{fig:threshold} and similar to the ones found in \cite{Baylam_2018}. Direct and inverse cross relaxation is taken into account through the parameters $P_{41}$ and $P_{22}$ respectively.

The systems of the Eqs. (\ref{Eq.rate_1}-\ref{Eq.rate_4}) can be coupled with an optical cavity (modeled as Fabry-Perot resonator) by considering the equation obtained starting from \cite{Baylam_2018}
\begin{equation}
\frac{\partial I_L \left ( t\right )}{\partial t}=2 \left ( \alpha_L \left ( t \right ) l- \frac{T}{2}-L\right )\frac{I_L \left ( t \right )+I_s\left ( t \right )}{T_R},
    \label{eq:laser_coupling}
\end{equation}
where $\alpha_L=\left (\sigma_{eL}N_2-\sigma_{aL}N_1 \right )$  is the amplification coefficient, \textit{l} is the medium thickness, $T=1-R_1$ is the output coupler transmission and $T_R=-2D/\big(c\ln[R_1 R_2 \left ( 1-L\right )^2]\big)$ is the cavity lifetime, where $R_1$, $R_2$ and \textit{D} are the  cavity mirrors reflectivity and cavity length respectively, while $L$ represents the combined residual cavity losses.
Eq.(\ref{eq:laser_coupling}) is initialized by  the  spontaneous emission intensity term    $I_s\left ( t \right )=N_2 \left(t \right )h \nu_L l \Omega/(4 \pi \tau_2)$, with $\Omega$  the smaller solid angle defined by the mirrors and $\nu_L$ the frequency of the emitted laser. The output laser intensity is given by $I_{out}\left (t\right)=TI_L\left (t \right)$.

 The total $\mathrm{Tm^{3+}}$ ion density with a  doping percentage of $\eta_d$ (at.\%) is given by $N\left (\eta_d \right)=2.8\eta_d\times\SI{e28}{\metre^{-3}}$ where, in our case, we have $\eta_d=0.04$  ($N=1.12\times\SI{e27}{\metre^{-3}}$).  According to \cite{Albalawi_2017,Gebavi_2010}, we use $\tau_i \left ( \eta_d \right )=\tau_{i0}/\left ( 1+A_i\eta_d^2\right )$, where    $\tau_{20}=3.4\;\mathrm{ms}$ and  $\tau_{40}=0.6\;\mathrm{ms}$ \cite{Albalawi_2017,Loiko_2018,Baylam_2018}.
 In our experimental conditions  $\tau_2\left ( 4\%\right)\simeq\SI{1.22}{\milli\second}$ and $\tau_4\left (4\% \right )\simeq\SI{63}{\micro\second}$, while $\tau_3\approx\SI{2}{\micro\second}$ is assumed to be independent of the dopant concentration.
 The branching ratio coefficients are given by $\beta_{31}=0.9793$, $\beta_{32}=0.0207$, $\beta_{41}=0.9035$, $\beta_{42}=0.0762$ and $\beta_{43}=0.0203$ \cite{Antipov_2011}. The cross relaxation mechanism is  dominated by the direct one $P_{22}/P_{41}=0.03-0.08$ \cite{1997_Falconieri,Albalawi_2017,Tao_2013}. Following Ref.\cite{Tao_2013}, the coefficient is  given by $P_{41}\left (\eta_d \right )=B\eta_d^2/\left ( \eta_d^2+\eta_0^2\right )$, where $\eta_0=4.3$~at.\% \cite{1996_Rustad} is the characteristic dopant concentration and $B=\SI{2.8e-22}{\cubic\metre\per\second}$ is obtained from \cite{1997_Falconieri} and \cite{Tao_2013}. We find $P_{41}\left( 4\%\right)=\SI{6.28e-29}{\cubic\metre\per\micro\second}$, with $\left (P_{41}\left( 4\%\right)N\left( 4\%\right) \right )^{-1}\simeq\SI{14.2}{\micro\second}$. The cross-relaxation parameter $\eta\ped{CR}$ can be numerically evaluated by considering the ratio $\eta\ped{CR}\left ( \peff\right )=N_2/N_{2}^{cr}$ under stationary conditions, where $N_{2}^{cr}$ are the results of the system of Eqs.(\ref{Eq.rate_1}-\ref{eq:laser_coupling}) when the cross-relaxation is forcefully turned off ($P_{41}=P_{22}\equiv0$).  Above a certain threshold, which is approximately given by the lasing threshold of the system with no CR,  the efficiency parameter becomes almost independent on the pump power and thus $\eta\ped{CR}\left ( \peff\right )\rightarrow \eta\ped{CR} $.

Numerical simulations are performed considering a perfect overlapping between pump  and laser waist, so that the  retrieved pump power actually corresponds to  the effective absorbed pump power \peff.   The slope efficiency  is  defined by $\eta\ped{sl}=P\ped{out}/(\peff-\peff^{th} )$, where $\peff{}^{th}$ is the effective pump threshold. This value is related to  $\eta\ped{CR}$  through  $\eta\ped{sl}\simeq\eta\ped{CR} \left( \lambda_P/\lambda_L\right)R_1/\left( R_1+L\right)$, where $\lambda_P$ and $\lambda_L$ are the pump and laser wavelength respectively. The limit is given by $\eta\ped{sl}\leq0.8$ for $\eta\ped{CR}=2$.  An alternative set of equations  is considered in Ref.\cite{Baylam_2018}, where $P_{41}=P_{22}\equiv0$, while the measured $\eta\ped{CR}$  is directly introduced  in Eq.(\ref{eq:laser_coupling}) by using   $\left (\eta\ped{CR}\sigma_{eL}N_2-\sigma_{aL}N_1 \right )$ instead of $\alpha_L$.  Simulation parameters are reported in Table (\ref{tab:my_label}).

\begin{table}[]
    \centering
    \begin{tabular}{cc}
    \toprule
    Symbol & Value \\
    \midrule
    $R_1$ & \num{0.97}  \\
    $R_2$ & \num{0.999}  \\
    $L$  & \num{0.013} \\
    $T_R$ &  \SI{0.40}{\nano\second} \\
    $\sigma_a$  & \SI{3.0d-25}{\square\metre}\\
    $\sigma_e$  & \SI{8.3d-26}{\square\metre}\\
    $\sigma_{aL}$  & \SI{8.4d-27}{\square\metre}\\
    $\sigma_{eL}$  & \SI{4.2d-25}{\square\metre}\\
    $w_0$ & \SI{145}{\micro\metre} \\
    $\Omega/4 \pi$ &  \num{0.0029} \\
    \bottomrule
    \end{tabular}
    \caption{Simulation parameters.}
    \label{tab:my_label}
\end{table}

\section{\label{sec:exp}Experimental setup}

\begin{figure}
\centering
\includegraphics[width=\columnwidth]{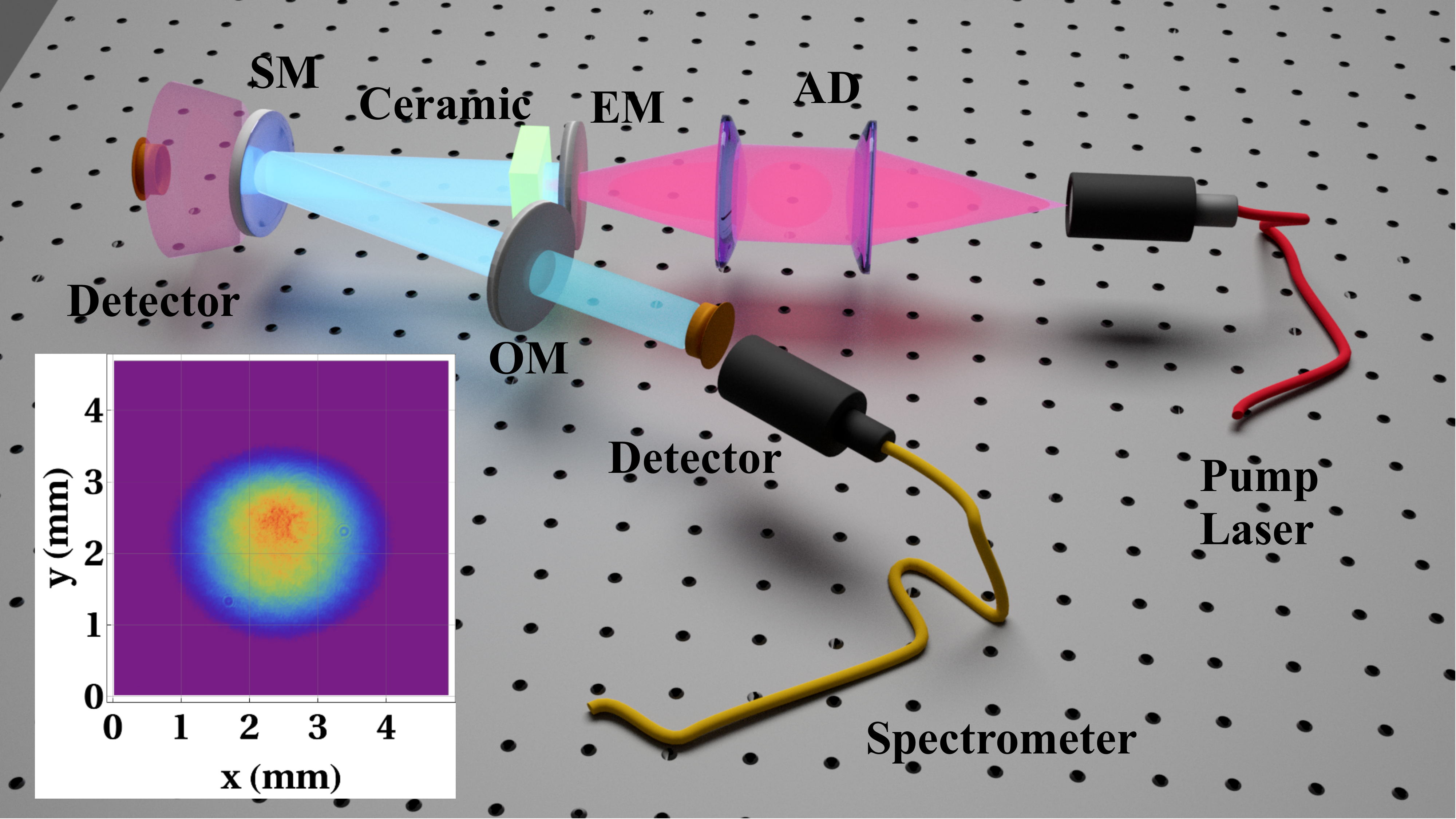}
\caption{Scheme (not to scale) of the experimental apparatus: the achromatic doublets (AD) are used to focus the pump beam emerging from the optical fiber on the sample; the cavity is composed by three mirrors: the dichroic entry mirror (EM) and the spherical mirror (SM) transmit the pump light while they reflect the \SI{2}{\micro\metre} radiation. 90\% and 97\% reflectivity output coupler mirrors (OM) are used. Both the pump and the laser beams are monitored in power and spectrum with photodiodes, power meters and spectrometers. In the inset the laser
spot captured at a distance of \SI{500}{\milli\metre} from the output coupler mirror with a Dataray WinCamD camera.}
\label{fig:scheme}
\end{figure}

The ceramic sample used in our study is $5\times 5\times 3.1$~\SI{}{\milli\cubic\metre} and is mounted on a water cooled
copper block. The thermal contact is ensured by a thin Indium foil
mounted between the lateral surface of the ceramic sample and the copper block itself.
We use two working temperature of \SI{13}{} and \SI{23}{\celsius} and the temperature is monitored by 2 1-wire sensors mounted on the sample holder.\\
A series of measurements of the absorption cross section as a function of the pump laser wavelength were performed in different positions of the sample, and revealed excellent sample homogeneity. An example
of such measure is reported in \fref{fig:sigmaabs} in good agreement with those reported in the literature \cite{Baylam_2018,Loiko_2018}.\\
Our test laser cavity is composed by the three mirror folded cavity depicted in \fref{fig:scheme}; in this configuration, optical pumping occurs along the longitudinal direction, in a similar configuration
as in \cite{Feng_2020}.
The pumping beam is obtained by a laser diode emitting at a measured wavelength of \SI{794.6 \pm 0.4}{\nano\metre}, hence in proximity of the absorption peak in \fref{fig:sigmaabs}.

 The laser diode is coupled to a multimode optical fiber of \SI{200}{\micro\metre} in diameter and a numerical aperture of 0.22. The beam emerging from the optical
fiber is focused by two piano-convex $f=\SI{50}{\milli\metre}$ optical doublets arranged in a $4f$ scheme.
We measure the pump beam spot size as a function of the distance
using a digital camera and the pump beam profile is well described
by a Gaussian function \cite{Moulton_1985,Siegman_1998} with
a waist of \SI{160}{\micro\metre} and an estimated $M^2=150$.
\begin{figure}
\centering
\includegraphics[width=0.9\columnwidth]{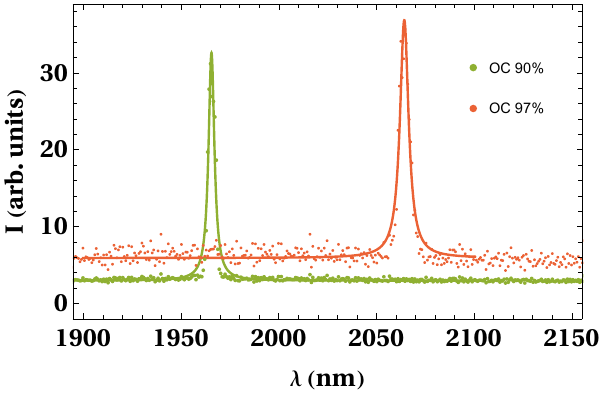}
\caption{Laser spectra for the two 90\% and 97\% reflectivity output coupler mirrors. With the 97\%
reflectivity we observe a change in the emission spectra as a function of the cavity losses due to its alignment.}
\label{fig:spectra}
\end{figure}
We point out that, given the full width half maximum of \SI{2}{\nano\metre}
of the pump diode laser emission we did not observe any significant change in the behavior
of the \tulio{} due to the slight detuning of the pump laser with respect to the exact absorption peak of \SI{796.2}{\nano\metre}, except for a slight decrease in the radiation absorbed by the sample
that is taken into account by the measurement procedure
we used, that is described below.

We operate the pump laser with pulses lasting \SI{10}{\milli\second}  repeated at a frequency of \SI{10}{\hertz} and we observe the laser emission peaked at both \SI{1965}{} and \SI{2065}{\nano\metre} as predicted
in \cite{Koopman_2011,Loiko_2018} and observed in \cite{Koopman_2011,Baylam_2018}. In our case we observe,  only the \SI{1965}{\nano\metre} wavelength emission when using the 90\% reflectivity output coupler while using the 97\%
reflectivity output coupler both
emission wavelengths or just the \SI{2065}{\nano\metre} one are visible, depending on the alignment of the cavity hence on the cavity losses, see  \fref{fig:spectra}. All the data that are presented
below in the text are taken with a well aligned laser emitting only at
\SI{2065}{\nano\metre}.

\section{\label{sec:expres}Experimental results}

We calculate the cavity
beam size by means of the ABCD formalism \cite{Kogelnik_1966} obtaining a waist of \SI{144\pm 5}{\micro\metre} roughly constant across the ceramic sample.

\begin{figure}[h]
\centering
\includegraphics[width=0.9\columnwidth]{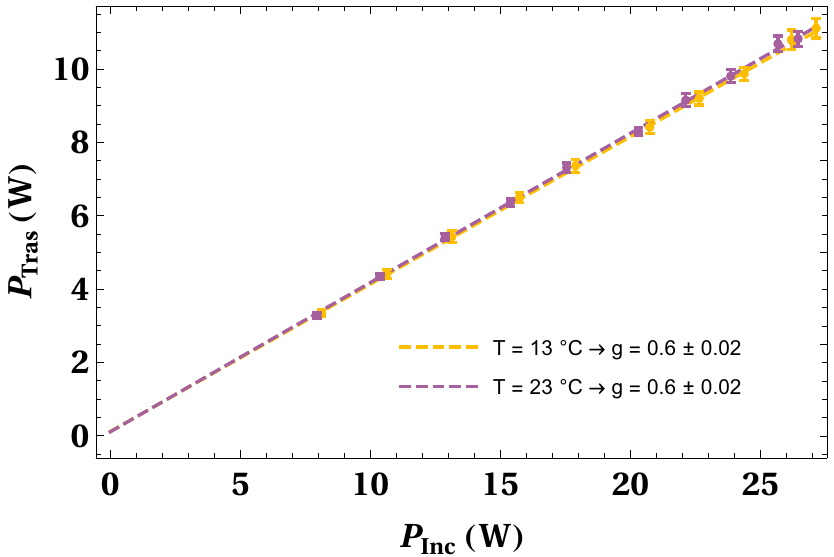}
\caption{Pump laser power transmitted as a function of the incident pump laser power for the two different working temperatures of \num{13} and \SI{23}{\celsius}. Straight lines are the results of a best fit calculation that we use to obtain the pump absorption ratio $g$ in Eq.~\eqref{eq:ptras}.}
\label{fig:ptras}
\end{figure}

Using the measured pump beam spot size and the laser beam model
we can calculate the average pump rate $\langle R\ped{P}\rangle$ considering the
volumetric overlap between the two beams in the lasing medium \cite{Svelto} as
\begin{equation}
\langle R\ped{p}\rangle=\frac{\alpha P\ped{Inc}}{h\nu\ped{p}}
\frac{
\int_0^d\frac{w\ped{L}(z)^2}{w\ped{L}(z)^2+w\ped{p}(z)^2}\mathrm{e}^{-\alpha z}\mathrm{dz}}{\frac{\pi}{2}
\int_0^d w\ped{L}(z)^2\mathrm{dz}}
\label{eq:peff}
\end{equation}
where $P\ped{Inc}$ is the pump laser power inside the sample.
Assuming the laser beam having a constant spot size $w\ped{L}$
inside the ceramic sample of length $d$
we can calculate the effective absorbed pump laser power \peff{} as
\begin{equation}
\label{eq:peff2}
\peff=\langle R\ped{p}\rangle\, \pi w\ped{L}^2 d\,  =\chi P\ped{Inc}.
\end{equation}
that in our case results in $\chi=\num{0.47\pm 0.06}$.

It is worth noting at this point the main limitations of the above procedure.
First, the laser beam profile inside the cavity is retrieved by the ABCD simulation; although this is a well consolidated procedure in such a kind of experiments (see for instance \cite{Loiko_2018,Baylam_2018}), it can result in some uncertainty.
Second, Eq.~\eqref{eq:peff2} holds in the case of a negligible depletion of
the ground state of the medium, since the absorption coefficient considered could
otherwise vary during laser operation.
For this reason we calculate the absorbed pump power as the difference between the
incident and the transmitted power right before and after the active medium respectively. A direct measurement of the transmitted power using a power meter after the active medium would stop the lasing condition making the measurement inconsistent.
Instead, we use a silicon photodiode to monitor the pump beam transmitted by the spherical cavity mirror with and without
the lasing condition. For the latter case we obstruct the light path between the spherical mirror and the output coupler in \fref{fig:scheme}. Finally
we obtain the absorbed pump power as,
\begin{equation}
    P\ped{Abs}=P\ped{Inc}-P\ped{Tras}\frac{V\ped{on}}{V\ped{off}} ,
\end{equation}
where $V\ped{on}$ and $V\ped{off}$ refer to the photodiode signal with and without lasing respectively, and considering as well the reflections at the ceramic faces.
With this procedure, we observe a difference up to the 10\% in
the transmitted pump power with or without lasing.
From the linear best fit to the data shown in \fref{fig:ptras} we can
observe that we always work below the saturation intensity hence
the absorbed pump laser power can be written as
\begin{equation}\label{eq:ptras}
P\ped{Tras}=(1-g) P\ped{Inc}
\end{equation}
resulting in $g=\num{0.6\pm0.02}$ in accordance
with the fraction calculated using the small signal absorption coefficient   $\alpha=\SI{322}{\metre^{-1}}$.

It should be noted that the value $P\ped{Abs}$ we obtain with this procedure gives us the
pump radiation absorbed over the whole volume described by the pump beam which is larger
than the laser beam, therefore we calculate $\peff=(\chi/g)P\ped{Abs}$
which matches \eqref{eq:peff2} with the substitution $P\ped{Abs}=gP\ped{Inc}$.

\begin{figure}
\centering
\includegraphics[width=0.9\columnwidth]{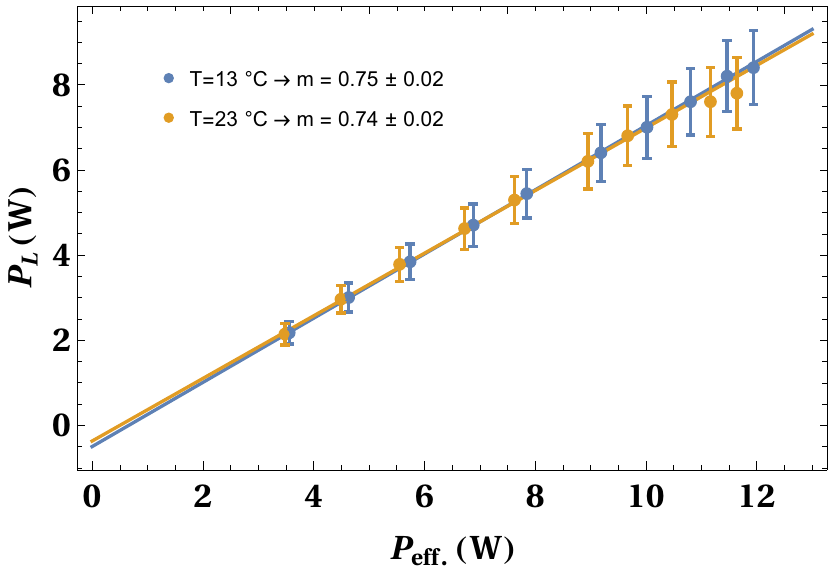}
\caption{Laser power as a function of the effective absorbed pump power for the two working temperature of \num{13} and \SI{23}{\celsius}; the straight lines are the results of a best fit calculation that provide  both the laser threshold power and slope efficiency.}
\label{fig:slope}
\end{figure}

The measured laser power as a function of \peff{} for the two working
temperatures reported in \fref{fig:slope} show a very small difference between
the two data set.

With our definition of \peff{} we obtain the slope efficiency of the laser
by means of a best fit calculation of the experimental data in Fig. \ref{fig:slope} with the equation
$P_L=m(\peff{}-\pthr)$ resulting in
$m_{13}=\num{0.75\pm0.02}$ and $m_{23}=\num{0.74\pm0.02}$, that gives a cross relaxation parameter
$\eta_{CR}=(\lambda_L / \lambda_{P}) \times m$
of \num{1.96\pm0.05} and \num{1.91\pm0.05} respectively at \num{13} and
\SI{23}{\celsius}.
We point out that these values are in agreement with values reported in recent literature\cite{Loiko_2018, Baylam_2018}, provided that the higher doping level of our sample and
the dependence of $\eta_{CR}$ upon the Tm concentration given in \cite{Zheng_2023} are taken into account.
\begin{figure}
\centering
\includegraphics[width=0.9\columnwidth]{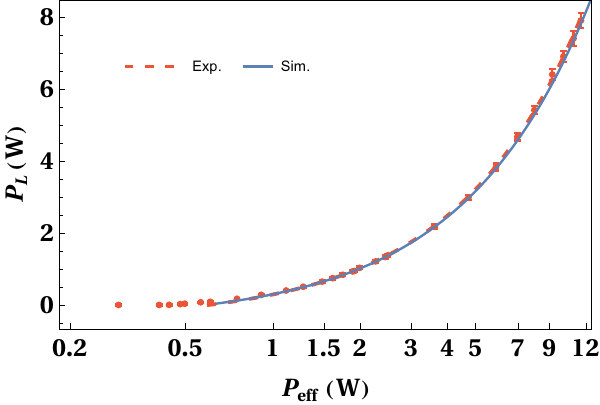}
\caption{Experimental and theoretical laser power as a function of
\peff{}. The dashed line is a linear fit of the data while the solid line is obtained with the model in Eq.~\ref{Eq.rate_1}---\ref{eq:laser_coupling}. The cavity energy loss $L$ is tailored to 1.3\% in order for the model to match the data.}
\label{fig:threshold}
\end{figure}
In order to measure the laser threshold power we added to
the apparatus a neutral density filter between the two
lenses of the pump beam optics to obtain lower
pump laser power. The resulting laser power is reported in \fref{fig:threshold}
with the same best fit calculation resulting in a laser threshold of $\pthr=\SI{0.6\pm0.02}{\watt}$
and in a slope efficiency $m=\num{0.73\pm0.02}$ and a cross relaxation coefficient of $\eta_{CR}=\num{1.89\pm0.05}$ with a working temperature of \SI{23}{\celsius}.
Data in \fref{fig:threshold} are superimposed on the numerical simulation
performed using the model described in sec.~\ref{sec:model} where
the parameter $L=1.1\%$ result from the best fit on our data. From the simulation it results a laser threshold of $\pthr=\SI{0.64}{\watt}$ and a cross-relaxation
parameter $\eta_{CR}=1.91$ in agreement with the experimental data.

\begin{figure}
\centering
\includegraphics[width=0.9\columnwidth]{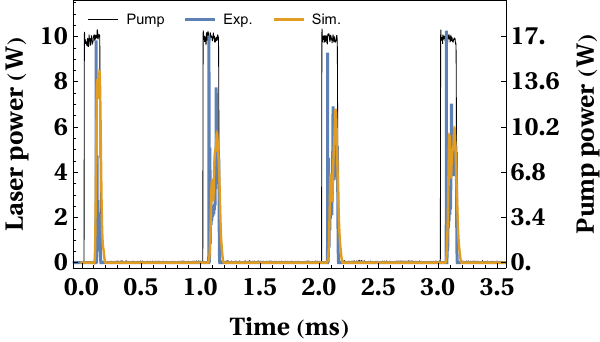}\\
\includegraphics[width=0.9\columnwidth]{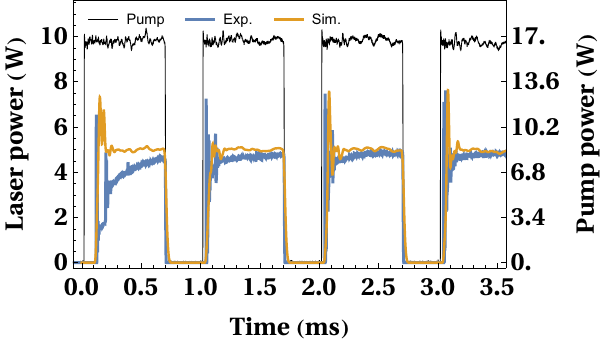}
\caption{Experimental and theoretical laser power as a function
of time obtained for pump pulse width of \SI{150}{\micro\second} in the top panel and \SI{700}{\micro\second} in the bottom panel.}
\label{fig:pulses}
\end{figure}

To further validate the theoretical model we simulated
the pulsed behavior of our system and compare the numerical results with the experimental one obtained by modulating the amplitude of the pump laser with
rectangular pulses of tunable time duration at a fixed
repetition frequency of \SI{1}{\kilo\hertz}. The data
obtained are shown in Fig.~\ref{fig:pulses} superimposed to the
simulation obtained using for the term $W_{41}$ in
Eq.~\ref{Eq.rate_1}---\ref{Eq.rate_4} the pump
waveform recorded with a power calibrated silicon photodiode. The laser power is recorded with a
InGaAs photodiode and the signal obtained is scaled
using the experimental slope efficiency of the laser.
The raw data are filtered with a numerical
low-pass filter with the cutoff frequency set
at the sampling rate of the oscilloscope used to record
the signals.
In this case the pulsed dynamics of the laser intensity over a millisecond timescale
is well reproduced by the simulation.
As a matter of fact, the exact temporal dynamics at the rising edge of the pulse depends critically on the actual pump laser intensity rising profile, whose behavior over $\sim\!\!10\,\mathrm{\mu s}$ timescales is not perfectly captured by our experimental apparatus.
Moreover, a rather complex emission dynamics, possibly involving both the emission wavelengths reported above, was also experimentally observed in \cite{Baylam_2018}; we're not accounting for this short timescale behavior, as it doesn't affect our comparison with the experimental data over the millisecond timescale considered in this paper.
It is worth
noticing that the delay of $\simeq\SI{100}{\micro\second}$ between the pump laser
rising wavefront and the laser emission in the first pulse is independent of the pulse duration while in the subsequent pulses the delay is shorter but depends on
time between the pulses.
As a further remark, again from \fref{fig:pulses} it can be noted that the laser emission amplitude
reaches the steady state within the few initial pulses, i.e., in a time scale comparable with the fluorescent time of the laser excited state $\tau_{40}$.

Finally, we can use the
the lifetime $\tau_{40}$ of the excited manifold  ${}^3\mathrm{H}_4$,  the $P_{41}$ coefficient, and Thulium concentration $N$ to
calculate the cross-relaxation coefficient. As reported\cite{2014_Dalfsen,Loiko_2018}, $\eta_{CR}=\frac{P_{41}N}{1/\tau_{40}+P_{41}N}$ and with our parameters,
it results in $\eta_{CR}=1.97$ in good agreement with our experimental and simulated values as well as with the values reported in literature.

\section{Conclusion}

We investigated the lasing operations and characteristics of a sample of \tulio{} with 4~at.\%{} doping ceramic using a three-mirror test optical cavity. The measure of the laser efficiency of 73\% at room temperature corresponds to a  cross-relaxation coefficient of $\sim$1.9 is in good agreement with the calculated value of 1.97.
A numerical model of the laser obtained solving
the macroscopic rate equations is presented and it reproduces with high accuracy the output laser power in both continuous and pulsed pumping regime.

\section{\label{sec:ackn}Acknowledgements}
The authors acknowledge the following funding: EU Horizon 2020 Research and Innovation Program EuPRAXIA Preparatory Phase, under Grant Agreement No. 101079773, EU Horizon IFAST, under Grant Agreement No.101004730. This research has been co-funded by the European Union - NextGeneration EU ``Integrated infrastructure initiative in Photonic and Quantum Sciences'' - I-PHOQS (IR0000016, ID D2B8D520, CUP B53C22001750006) and ``EuPRAXIA Advanced Photon Sources'' - EuAPS (IR0000030, CUP I93C21000160006). We also acknowledge contribution to the strategic view of this research by the Project “Tuscany Health Ecosystem—THE” “Spoke 1—Advanced Radiotherapies and Diagnostics in Oncology” funded by the NextGenerationEU (PNRR), Codice progetto ECS00000017, D.D. MUR No. 1055 23 May 2022.

\end{document}